\documentstyle[11pt,epsf,rotate,newpasp,twoside]{article}
\newcommand{\kms}{$\mbox{km~s}^{-1}$}
\newcommand{\myr}{$\mbox{mas~yr}^{-1}$}

\def\edcomment#1{\iffalse\marginpar{\raggedright\sl#1\/}\else\relax\fi}
\reversemarginpar

\begin{document}
\title{Measurement of the proper motion of Vela with the LBA.}
 \author{Richard Dodson}
\affil{ISAS, Sagamihara, Kanagawa, Japan, 229-8510}
\author{David Legge, Peter McCulloch, John Reynolds}
\affil{University of Tasmania, Hobart; ATNF, CSIRO, Australia}

\vspace{-0.3cm}
\begin{abstract}
The first accurate measurements of Vela's proper motion were produced
using the Parkes-Tidbinbilla Interferometer (PTI) (Bailes et al
1989). This 275 km baseline at 1.6~GHz gave a resolution of 140 mas
and detected the proper motion and thus, by extrapolation, the birth
site. We have extended this work to allow the third dimension to be
derived via the parallax. We find the offset between the spin axis and
the proper motion vector to be $9^o$, which limits acceleration time
from the SN blast.
\end{abstract}

\vspace{-1.0cm}
\section*{Introduction}
\vspace{-0.3cm}
Our results allow the testing of the prediction of
\cite{spruit_phinney} on the alignment of the spin axes of pulsars
with their proper motion vector. The spin axis can be derived from the
high resolution images from Chandra. These allow the symmetry, and
thus presumably the spin, axes to be directly discerned. Only two
cases have been accurately tested so far, the Crab (\cite{crab_hst}) and
Vela. The extra accuracy we can provide refines the conclusion reached
by Pavlov et al (2000) and Helfand et al (2001) for the Vela
alignment.
If the model of \cite{spruit_phinney} is correct any misalignment of
the axes allows the us to estimate the timescale of the impulse in
terms of the initial rotation period, a value of great importance to
understanding the core collapse. The rotations of the pulsar average
out the off-axis acceleration of the pulsar, bringing the proper
motion and the spin axis into alignment.

\vspace{-0.6cm}
\section*{Results}
\vspace{-0.4cm}
The data, and its reduction are full described in Dodson et al (2003)
We find a pulsar position of $\alpha_{J2000} = 08^h35^m20^s.61149 \pm
0.00002$, $\delta_{J2000} = -45^\circ10^\prime34^{\prime \prime}.8751
\pm 0.0003$ for a reference epoch of 2000.0. The proper motion is
$\mu_{\alpha {\rm cos}\delta} = -49.68 \pm 0.06,\ \mu_\delta = 29.9
\pm 0.1$ \myr\ and a parallax of $3.5 \pm 0.2$ mas, or
$287_{-17}^{+19}$ pc.

Two corrections need to be applied to our results to calculate the
motion of the Vela pulsar in its local environment. Our observations
are directly tied to the ICRF, therefore the solar peculiar motion and
the galactic rotation contribute to the observed proper motion, and
need to be removed. The dominant source of error is from the
uncertainties in the solar peculiar motion parameters. We have used
the measured uncertainties in our observations and combined those with
the models. This gives us an angular motion, at Vela's local standard
of rest, of $\mu_\alpha = -38.6 \pm 1.2$ \myr , $\mu_\delta = 23 \pm
1.5$ \myr\ or $\mu_* = 45\pm 1.3$ \myr\ at a position angle of
$301^\circ \pm 1.8$.  The pulsars' transverse space velocity is
therefore $61 \pm 2$ \kms . The PA does not lie exactly along the spin
axis (e.g. $310^\circ \pm 1.5$ (Helfand et al 2001), or $307^\circ \pm
2$ (Pavlov et al 2001)), which implies that the impulse timescale
was not quite long enough to average the off-axis component to zero.

Recent communications with Caraveo {\em et al} has found that the
published HST position angle (Caraveo etal 2001) has an arithmetic
error and is in fact $302.2 \pm 1.6$, in excellent agreement with our
own result. Thus we consider the implications of this offset.

Firstly if the impulse is equal along and across the spin axis,
i.e. at $45^o$, after only two rotations it is no longer possible to
achieve an offset of $9^o$. If the acceleration vector was at $81^o$
to the spin axis it is still possible to generate such an offset until
the seventh rotation. This limits either the initial rotation period,
or the timescale of the SN event.


\vspace{-0.6cm}
\section*{Conclusions}
\vspace{-0.4cm}

We have measured the proper motion and parallax of the Vela pulsar to
an unprecedented accuracy ($\mu_{\alpha {\rm cos}\delta}= -49.68 \pm
0.06,\ \mu_\delta= 29.9 \pm 0.1$ \myr\ , $\pi = 3.5 \pm 0.2$ mas), and
have been able to convert these back to the space velocity and
position angle of the pulsar in its local environment with greater
precision that previously possible ($61 \pm 2$ \kms at
$301^\circ\pm1.8$), because of the unambiguity in the radio reference
frame. This allows the precise comparison of the Vela X-ray nebula
symmetry axis and the proper motion of the Vela pulsar, opening
insights into the timescale of the core collapse processes. We
estimate that the collapse must have taken less than seven rotations
to occur.

\begin{figure}
\plottwo{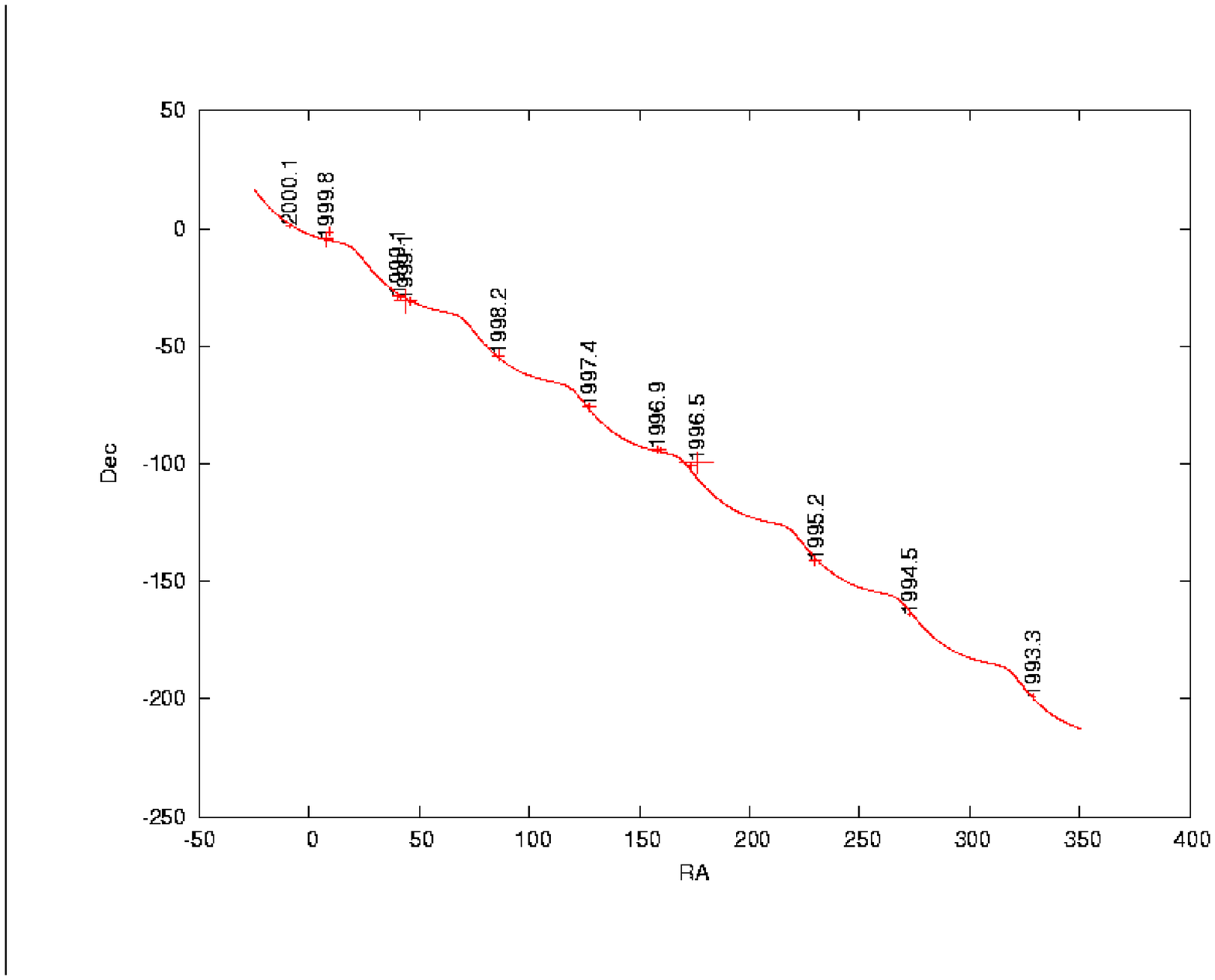}{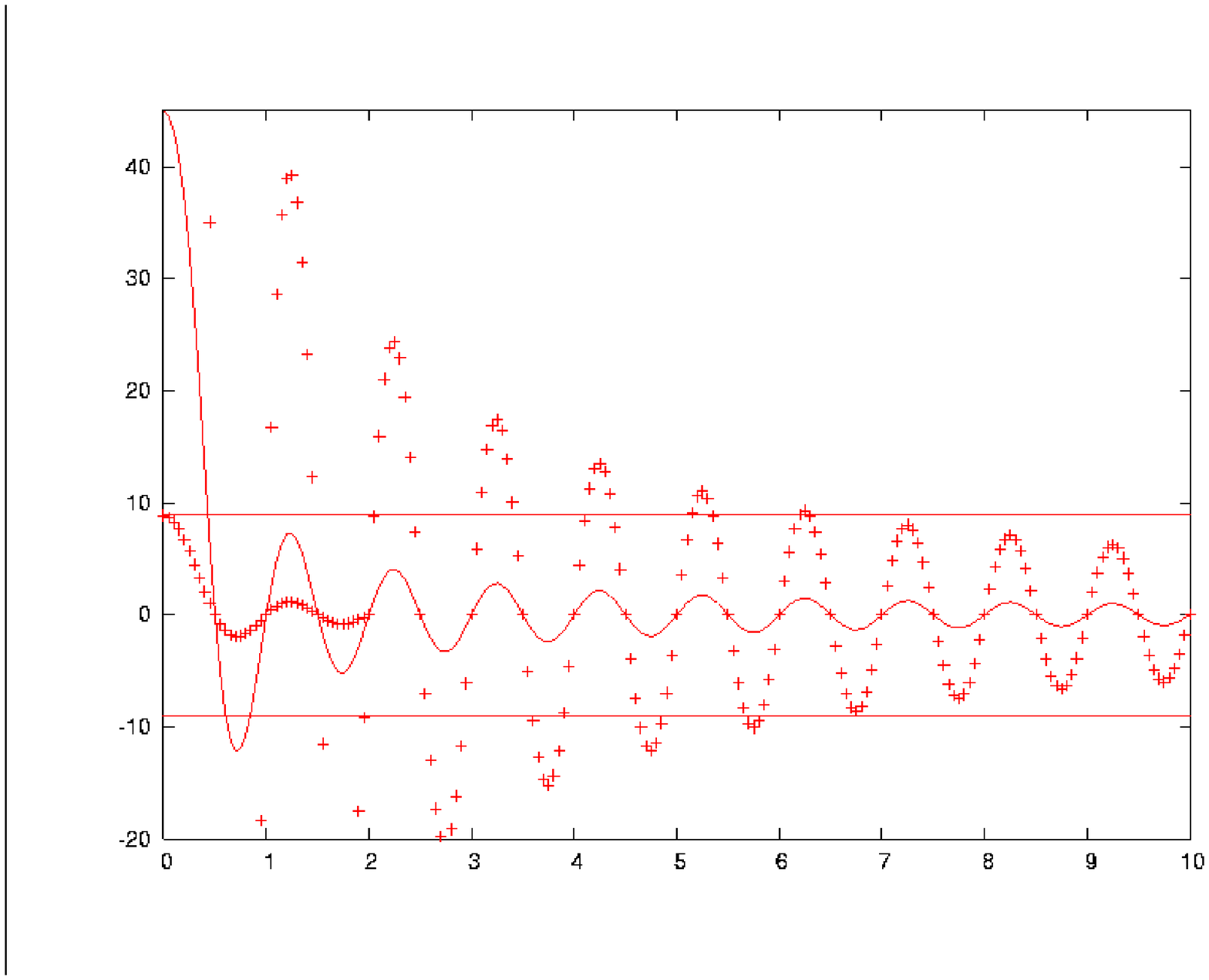}
\caption{a) The position of the Vela pulsar over the duration of the
experiment, the epochs and the errors are marked. b) The offset
between the projected proper motion and the spin axes for an impulse
at $45^o$ (solid) line, at $9^o$ to the spin axis (close spaced dots)
and at $81^o$ to the spin axis (widely spaced dots).}
\label{fig:onsky}  
\end{figure}

\begin{footnotesize}

\end{footnotesize}
\end{document}